# Fog Computing& IoT: Overview, Architecture and Applications

## Harshit Gupta[1], Dr. Ajay Kumar Bharti[2]

M.Tech.,Dept. of Computer Science, Maharishi University of Information Technology, Lucknow, India[1]

Dr. Ajay Kumar Bharti, Dept. of Computer Science, Maharishi University of Information Technology, Lucknow, India[2]

**Abstract**: Fog computing is an emerging technology in the field of network services where data transfer from one device to another to perform some kind of activity. Fog computing is an extended concept of cloud computing. It works in-between the Internet of Things (IoT) and cloud data centers and reduces the communication gaps. Fog computing has made possible to have decreased latency and low network congestion. Fog computing is an on-going research trend in which the possibility of efficient network services exist. Fog computing can be described as a cloud type platform having similar services of data computation, data storage and application service but it is fundamentally different as it decentralized. In this paper, we have done a comprehensive survey on fog computing& IoT and described the fog computing architecture and analyse its different benefits and applications. We have also analysed the security aspects of fog computing & IoT, which is necessary and an important part of any kind of technology used in data communication system.

## I. INTRODUCTION

The Internet of things (IoT) will be the Internet of future, as we have seen a huge increase in wearable technology, smart grid, smart home/city, smart connected vehicles. International Data Corporation (IDC) has predicted that in the year of 2015, "the IoT will continue to rapidly expand the traditional IT industry" up 14% from 2014 [1]. Since smart devices are usually inadequate in computation power, battery, storage and bandwidth, IoT applications and services are usually backed up by strong server ends, which are mostly deployed in the cloud, since cloud computing is considered as a promising solution to deliver services to end users and provide applications with elastic resources at low cost. However, cloud computing cannot solve all problems due to its own drawbacks. Applications, such as real time gaming, augmented reality and real time streaming, are too latency-sensitive to deploy on cloud. Since data centers of clouds are located near the core network, those applications and services will suffer unacceptable round-trip latency, when data are transmitted from/to end devices to/from the cloud data center through multiple gateways. Besides this, there are also problems unsolved in IoT applications that usually require mobility support, geo-distribution and location-awareness.

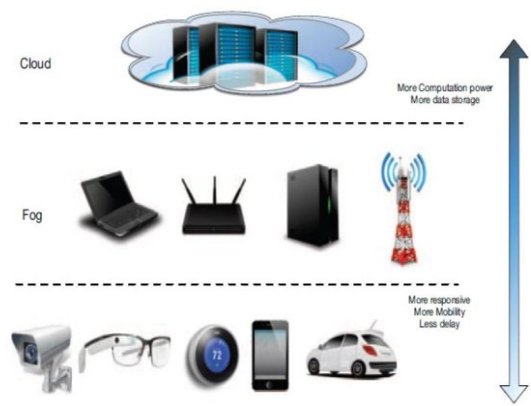

Fig. 1.Three layer architecture: end user/fog/cloud

Fog computing is usually cooperated with cloud computing. As a result, end users, fog and cloud together form a three layer service delivery model, as shown in Fig. 1[20]. Fog computing also shows a strong connection to cloud computing in terms of characterization. For example, elastic resources (computation, storage and networking) are the building blocks of both of them, indicating that most cloud computing technologies can be directly applied to fog computing. However, fog computing has several unique properties that distinguish it from other existing computing architectures. The most important is its close distance to end users. It is vital to keep computing resource at the edge of the network to support latency-sensitive applications and services. Another interesting property is location-awareness; the geo-distributed fog node is able to infer its own location and track end user devices to support mobility. Finally, in the era of big data, fog computing can support edge analytics and stream mining, which can process and reduce data volume at a very early stage, thus cut down delay and save bandwidth.

In the paper, we focus on the fog computing platform design and applications. We will briefly review existing platforms and discuss important requirements and design goals for a standard fog computing platform. We will also introduce some IoT applications to promote the fog computing.

## II. FOG COMPUTING OVERVIEW

**Definition**: There are a few terms similar to fog computing, such as mobile cloud computing, mobile edge computing, etc. Below we explain each of them.

1) **Local Cloud:** Local cloud is a cloud built in a local network. It consists of cloud-enabling software running on local servers and mostly supports interplay with remote cloud. Local cloud is complementary to remote cloud by running dedicated services locally to enhance the control of data privacy.
2) **Cloudlet:** Cloudlet is "a data center in a box", which follows cloud computing paradigm in a more concentrated manner and relies on high-volume servers [4]. Cloudlet focuses more on providing services to delay-sensitive, bandwidth-limited applications in vicinity.
3) **Mobile Edge Computing:** Mobile edge computing [5] is very similar to Cloudlet except that it is primarily located in mobile base stations.
4) **Fog Computing:** Fog computing is a geographically distributed computing architecture with a resource pool consists of one or more ubiquitously connected heterogeneous devices (including edge devices) at the edge of network and not exclusively seamlessly backed by cloud services, to collaboratively provide elastic computation, storage and communication (and many other new services and tasks) in isolated environments to a large scale of clients in proximity..

## III. BENEFITS OF FOG COMPUTING

Fog computing expands the cloud computing model to the edge of the network. Although the fog and the cloud use similar resources (networking, computing and storage) and share many of the same mechanisms and attributes (virtualization, multi-tenancy), fog computing brings many benefits for IoT devices. These benefits can be summarized as follows:

- Greater business agility: With the use of the right tools, fog computing applications can be quickly developed and deployed. In addition, these applications can program the machine to work according to the customer needs.
- Low latency: The fog has the ability to support real-time services (e.g., gaming, video streaming)
- Geographical and large-scale distribution: Fog computing can provide distributed computing and storage resources to large and widely distributed applications
- Lower operating expense: Saving network bandwidth by processing selected data locally instead of sending them to the cloud for analysis
- Flexibility and heterogeneity: Fog computing allows the collaboration of different physical environments and infrastructures among multiple services.

## IV. FOG ARCHITECTURE

Architecture of fog computingrepresented in the Fig (2)[16].Typically, the architecture of IoT is divided into three basic layers : (i) application layer, (ii) network layer, and (iii) perception layer, which are further described below.

**(i)** *Perception layer*, also known as the sensor layer, is implemented as the bottom layer in IoT architecture [11]. The perception layer interacts with physical devices and components through smart devices (RFID, sensors, actuators, etc.). Its main objectives are to connect things into IoT network, and to measure, collect, and process the state information associated with these things via deployed smart devices, transmitting the processed information into upper layer via layer interfaces.

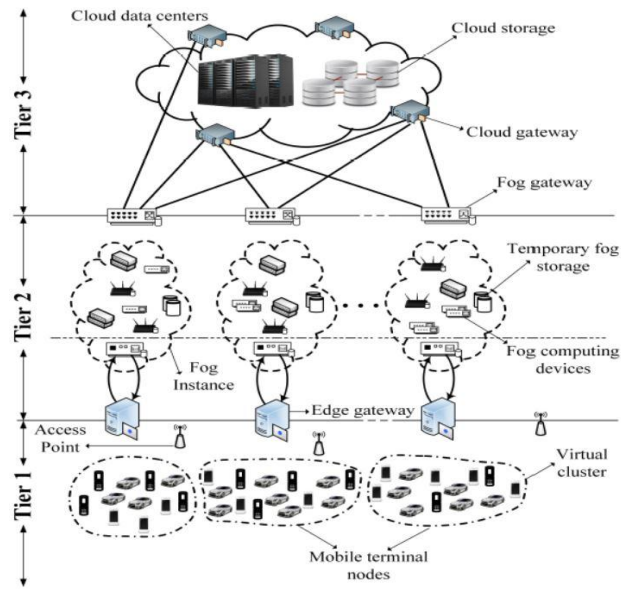

Fig. 2. Fog Architecture

**(ii)** *Network layer*, also known as the transmission layer, is implemented as the middle layer in IoT architecture. The network layer is used to receive the processed information provided by perception layer and determine the routes to transmit the data and information to the IoT hub, devices, and applications via integrated networks. The network layer is the most important layer in IoT architecture, because various devices (hub, switching, gateway, cloud computing perform, etc.), and various communication technologies (Bluetooth, WiFi, Long-Term Evolution (LTE), etc.) are integrated in this layer. The network layer should transmit data to or from different things or applications, through interfaces or gateways among heterogeneous networks, and using various communication technologies and protocols.

**(iii)** *Application layer*, also known as the business layer, is implemented as the top layer in IoT architecture [7]. The application layer receives the data transmitted from network layer and uses the data to provide required services or operations.For instance, the application layer can provide the storage service to backup received data into a database, or provide the analysis service to evaluate the received data for predicting the future state of physical devices. A number of applications exist in this layer, each having different requirements. Examples include smart grid, smart transportation, smart cities, etc.

## V. SECURITY AND PRIVACY CHALLENGES IN IOT

Although the IoT can play a central role in delivering a rich portfolio of services more effectively and efficiently to end users, it could impose security and privacy challenges. In the following, we summarize the major security and privacy challenges in IoT environments.

**Authentication**
Authentication is an essential requirement for the security of IoT devices. Unfortunately, many IoT devices don't have enough memory and CPU power to execute the cryptographic operations required for an authentication protocol. These resource-constrained devices can outsource expensive computations and storage to a fog device that will execute the authentication protocol. Yee Wei Law and colleagues proposed a wide-area measurement system key management (WAKE) model for the smart grid. This model is based on Public-Key Infrastructure (PKI) using multicast authentication for secure communications. While traditional PKI based authentication could solve the problem, it wouldn't scale well for IoT systems.

**Trust**
Due to the nature of the IoT environment, which integrates various devices and sensors belonging to multiple actuators, the following question arises: To what degree can we trust the IoT devices? There's no efficient mechanism that can measure when and how to trust IoT devices. In the absence of a trust measurement, users of IoT services need to consider whether it's profitable to abstain from using certain IoT services. Therefore, cultivating the trust between IoT devices plays a central role in establishing secure environments to preserve the security and reliability of IoT services. Trust models based on reputation have been successfully deployed in many scenarios such as online social networks.

**Privacy**

The privacy leakage of user information in IoT environments, such as data, location, and usage, is attracting the attention of the research community. The resource-constrained IoT devices lack the ability to encrypt or decrypt generated data, which makes it vulnerable to an adversary. Another privacy issue is the location privacy that can be used to infer the IoT device's location. Several IoT applications are location based services, especially mobile computing applications. An adversary can infer the IoT device's location based on the communication patterns. The last privacy issue is the protection of a user's usage pattern of some generated data by IoT devices, such as in the smart grid. For instance, the readings of smart meters can reveal many usage patterns of IoT clients, such as how many people live in the household, when they turn on the TV, or when they are at home. Many privacy-preserving schemes have been proposed in different IoT applications such as smart grids, healthcare systems, and vehicle ad hoc networks. However, the resource-constrained IoT devices limit the techniques that can be used to deliver efficient and effective privacy-preserving schemes.

## VI. APPLICATION S OF FOG AND IOT

The fog computing platform has a broad range of applications. Bonomi et al. [2] have presented fog computing scenarios in connected vehicle, smart grid and Wireless Sensor And Actuator Networks (WSAN). Later, Stojmenovic et al. [8],have emphasised previous scenarios and expanded fog computing on smart building.

*A. Smart Home*

With the rapid development of the Internet of Things, more and more smart devices and sensors are connected at home. However, products from different vendors are hard to work together. Some tasks, which require large amount of computation and storage, , e.g. real-time video analytics, are infeasible due to the limited capability of hardware. To solve these problems, fog computing is utilized to integrate all debris into a single platform and empower those Smart Home applications with elastic resources.

To use home security application as an example, widely deployed secure sensors consist of smart lock, video/audio recorder, various sensor monitors (e.g. light sensor, occupancy sensor, and motion sensor etc). If not products of same vendor, those secure devices are hard to combine. A motion sensor detects a suspicious motion in a certain room, then a cleaning robot with video camera will be commanded to check out the exact location. Real-time video analytics will process those video and confirm whether it is a false alarm. Notification and All paragraphs must be indented. report will be sent to house owner and the system will call police if applicable.

*B. Smart Traffic Lights*

Fog computing allows traffic signals to open roads depending on sensing flashing lights. It senses the presence of pedestrians and cyclists and measures the distance and speed of the nearby vehicles. Sensor lighting turns on when it identifies movements and vice-versa [17]. Smart traffic lights may be considered to be fog nodes which are synchronized with each other to send warning messages to nearby vehicles. The interactions of the fog between the vehicle and access points are improved with WiFi, 3G, smart traffic lights and roadside units [7].

*C. Health Data Management*

Health data management has been a sensitive issue since health data contains valuable and private information. With fog computing, it is able to realize the goal that patient will take possession of their own health data locally. Those health data will be stored in fog node such as smartphone or smart vehicle. The computation will be outsourced in a private-preserving manner when patient is seeking help from a medical lab or a physician's office. Modification of data happens directly in patient-owned fog node.

## VII. CONCLUSION AND FUTURE WORK

We have briefly introduced fog computing and given a more comprehensive definition of fog computing and internet of thing after analysing various interrelated concepts. In this paper, we have discussed the architecture of fog computing for smooth functioning of internet of things, benefits and challenges of a fog platform. We have also presented the design and implementation of a prototyping platform for fog computing. There is also fog based application analysis like Smart Home applications, Smart traffic light etc. Future work includes comparative analysis of cloud and fog computing in context of power consumption, cost incurring in application process and latency in service; for Internet of Things. Also extending our work by implementing a prototype which will support real implementation of fog based internet of things.

## REFERENCES

[1] F. Bonomi, R. Milito, J. Zhu, and S. Addepalli, "Fog computing and its role in the internet of things," in *workshop on Mobile cloud computing*. ACM, 2012.

[2] S. Yi, Z. Qin, and Q. Li, "Security and privacy issues of fog computing: A survey," in *International Conference on Wireless Algorithms, Systemsand Applications (WASA)*, 2015.
[3] M. Satyanarayanan, P. Bahl, R. Caceres, and N. Davies, "The case for vm-based cloudlets in mobile computing," *Pervasive Computing*, 2009.
[4] ETSI, "Mobile-edge computing," http://goo.gl/7NwTLE, 2014, [Online; accessed 18-June-2015].
[5] H. T. Dinh, C. Lee, D. Niyato, and P. Wang, "A survey of mobile cloud computing: architecture, applications, and approaches," *WCMC*, 2013.
[6] L. M. Vaquero and L. Rodero-Merino, "Finding your way in the fog: Towards a comprehensive definition of fog computing," *ACMSIGCOMM CCR*, 2014.
[8] I. Stojmenovic, "Fog computing: A cloud to the ground support for smart things and machine-to-machine networks," in *TelecommunicationNetworks and Applications Conference (ATNAC)*. IEEE, 2014.
[9] I. Stojmenovic and S. Wen, "The fog computing paradigm: Scenarios and security issues," in *Federated Conference on Computer Science andInformation Systems (FedCSIS)*. IEEE, 2014.
[10] K. Hong, D. Lillethun, U. Ramachandran, B. Ottenwalder,¨ and B. Kold-ehofe, "Mobile fog: A programming model for large-scale applications on the internet of things," in *ACM SIGCOMM workshop on Mobilecloud computing*, 2013.
[11] J. Zhu *et al.*, "Improving web sites performance using edge servers in fog computing architecture," in *SOSE*. IEEE, 2013.
[12] H. Madsen, G. Albeanu, B. Burtschy, and F. Popentiu-Vladicescu, "Reliability in the utility computing era: Towards reliable fog comput-ing," in *IEEE International Conference on Systems, Signals and ImageProcessing (IWSSIP)*, 2013.
[13] K. Hong, D. Lillethun, U. Ramachandran, B. Ottenwalder,¨ and B. Kold-ehofe, "Opportunistic spatio-temporal event processing for mobile situation awareness," in *Proceedings of the ACM international conferenceon Distributed event-based systems*, 2013.
[14] B. Ottenwalder,¨ B. Koldehofe, K. Rothermel, and U. Ramachandran, "Migcep: operator migration for mobility driven distributed complex event processing," in *Proceedings of the ACM international conferenceon Distributed event-based systems*, 2013.
[15] S. Yi, C. Li, and Q. Li, "A survey of fog computing: Concepts, applications and issues," in *Proceedings of the 2015 Workshop on MobileBig Data*. ACM, 2015.
[16] SubhadeepSarkary, SubarnaChatterjee, SudipMisraz et al, Assessment of the Suitability of Fog Computing in the Context of Internet of Things, 2015
[17] J. K. Zao*et al.*, "Pervasive brain monitoring and data sharing based on multi-tier distributed computing and linked data technology," *Frontiersin human neuroscience*, 2014.
[18] Y. Shi, S. Abhilash, and K. Hwang, "Cloudlet mesh for securing mobile clouds from intrusions and network attacks," in *The Third IEEEInternational Conference on Mobile Cloud Computing, Services, and Engineering*, 2015.
[19] M. Satyanarayanan, Z. Chen, K. Ha, W. Hu, W. Richter, and P. Pil-lai, "Cloudlets: at the leading edge of mobile-cloud convergence," in *IEEE International Conference on Mobile Computing, Applications and Services (MobiCASE)*, 2014.
[20] Shanhe Yi, ZijiangHao, Zhengrui Qin, and Qun Li et al, Fog Computing: Platform and Applications, 2015.
[21] M. Satyanarayanan, R. Schuster, M. Ebling, G. Fettweis, H. Flinck, K. Joshi, and K. Sabnani, "An open ecosystem for mobile-cloud convergence," *IEEE Communications Magazine*, 2015.
[22] Y. Cao, P. Hou, D. Brown, J. Wang, and S. Chen, "Distributed analytics and edge intelligence: Pervasive health monitoring at the era of fog computing," in *Proceedings of the 2015 Workshop on Mobile Big Data*. ACM, 2015.
[23] M. A. Hassan, M. Xiao, Q. Wei, and S. Chen, "Help your mobile applications with fog computing," in *Fog Networking for 5G and IoTWorkshop*, 2015.
[24] K. Ha, Z. Chen, W. Hu, W. Richter, P. Pillai, and M. Satyanarayanan, "Towards wearable cognitive assistance," in *Mobisys*. ACM, 2014.
[25] Cisco, "Iox overview," http://goo.gl/n2mfiw, 2014, [Online; accessed 18-June-2015].

## BIOGRAPHIES

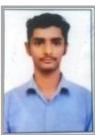

**Harshit Gupta**, M.Tech.Scholar in Department of Computer Science & Engineering at Maharishi University of Information Technology, Lucknow (U.P.) India. His research interest is in Fog computing and IoT and Artificial Intelligenc.

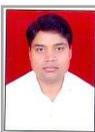

**Dr. Ajay Kumar Bharti**, Professor, Computer Science & Engineering at Maharishi University, Lucknow(U.P.) India. His research interest is e-Governance, Service Oriented Architecture and Knowledge Based System. He has published number of research papers in reputed journals and conferences.